\documentstyle[epsfig,12pt]{article}

\begin{document}
{\large\bf\sloppy\centering EFFECT OF CRACK BLUNTING ON SUBSEQUENT
  CRACK PROPAGATION\\[\baselineskip]

  }
J. SCHI{\O}TZ$^\ast$, A. E. CARLSSON$^\ast$, L. M.
CANEL$^\ast$, and ROBB THOMSON$^{\ast\ast}$\\
$^\ast$Department of Physics, Washington University, St. Louis,
MO 63130-4899\\
$^{\ast\ast}$National Institute of Standards and Technology, Gaithersburg,
MD 20899

\section{Abstract}

Theories of toughness of materials depend on an understanding of the
characteristic instabilities of the crack tip, and their possible
interactions.  In this paper we examine the effect of dislocation
emission on subsequent cleavage of a crack and on further dislocation
emission.  The work is an extension of the previously published
Lattice Greens Function methodology\cite{ThZhCaTe92,ZhCaTh94,CaCaTh95}.
We have developed a Cavity Greens Function describing a blunt crack
and used it to study the effect of crack blunting under a range of
different force laws.  As the crack is blunted, we find a small but
noticeable increase in the crack loading needed to propagate the
crack.  This effect may be of importance in materials where a
dislocation source near the crack tip in a brittle material causes the
crack to absorb anti-shielding dislocations, and thus cause a blunting
of the crack.  It is obviously also relevant to cracks in more
ductile materials where the crack itself may emit dislocations.

\section{Introduction}

When a sharp crack in a material is loaded until it deforms plastically
at the crack tip, two fundamentally different modes of
deformation can occur.  The crack may propagate (possibly leading to
cleavage of the specimen), or it may emit a dislocation.  In the first
case the material is said to be intrinsically brittle, in the second
it is intrinsically ductile.  Even for single crystals, many other
aspects of the micro-structure of the material influence the behavior
of the material, such as dislocation activity in the material
surrounding the crack and shielding of the stress fields by the
dislocations present in the vicinity of the crack.  In this paper we
will however concentrate on the intrinsic behavior of the crack.

The intrinsic behavior of the crack can be predicted by comparing the
critical stress intensity factors necessary for cleavage and for
dislocation emission.  The cleavage is well described by the Griffith
criterion\cite{Gr20}, but several criteria have been proposed for the
dislocation emission.  Rice\cite{Ri92} has proposed an emission criterion
given by a well-defined solid-state parameter, the unstable stacking
fault energy ($\gamma_{us}$), characterizing the barrier to
displacement along the slip plane.  In this model the emission
criterion becomes a ``balance'' between two energies: The surface
energy ($\gamma_s$) and the unstable stacking energy.  Zhou {\em et
  al.}\cite{ZhCaTh94,ZhCaTh93} extended this model to include the
energy of the small ledge at the end of the crack, and showed that for
most physically realistic force laws this term dominates the
energetics, leading to an emission criterion containing both
$\gamma_s$ and $\gamma_{us}$ and a ductility criterion
that is independent of $\gamma_s$.

These ductility criteria all assume that the crack is sharp at the
atomic level.  However, emission of a dislocation will change the
local geometry of the crack, possibly changing both the emission and
cleavage criteria.  Furthermore, a crack may {\em absorb\/} a
dislocation, coming from sources in the neighborhood of the crack tip.
This may also lead to blunting of the crack.  In the following we will
discuss the effects such blunting may have on the dislocation emission
and cleavage of the crack.

\section{Methodology}

Ordinary molecular dynamics (MD) simulations of cracks are very
difficult, due to the extremely slow decay of the stress fields
($\sigma \propto r^{-1/2}$).  This makes it necessary to use very
large systems to minimize the effects of boundary conditions; millions
of atoms may be required even for two-dimensional simulations.  On the
other hand all of the interesting phenomena will be concentrated near the
crack tip, and most of the atoms are just propagating the elastic
field.  This leads to a very inefficient use of computer resources.

Our solution is to model the elastic response of the surrounding media
by a Green's function.  The atoms are divided in two classes. The
atoms near the crack tip interact with each other through a non-linear
force law (the {\em non-linear zone\/}), whereas the atoms far from the
crack tip interact only through linear forces.  This {\em linear
  zone\/} can then be fully described by a lattice Green's function
$G_{ij}({\bf r}, {\bf r}')$, describing the response of the atom at
${\bf r}$ to a force acting on the atom at ${\bf r}'$.  One needs only
the Green's function elements involving the nonlinear zone and a
defect zone.  The Green's function can be calculated in a
computationally efficient way.  The procedure for calculating the
Green's function, and for introducing a defect (a crack in this case)
has been discussed by Zhou {\em et al.}\cite{ThZhCaTe92}.  The total
energy of the system can then be described as the sum of the energy in
the elastic far field (calculated from the Green's function) and in
the non-linear interactions.  In this way the total energy of the
system is described as a function of a much reduced number of degrees
of freedom (the positions of the $10^2$--$10^3$ atoms in the non-linear
zone, compared to the $10^6$--$10^7$ atoms in the full problem).  The
problem can then be treated with conventional minimization
techniques\cite{CaCaTh95}.

We use this method to study the deformation modes of blunt cracks
with up to seven atomic layers of blunting, for a range of force laws.
Figure \ref{fig:modes}a show a typical initial configuration.  We have
a region in front of the crack where the atoms are allowed to move
freely, and two spurs are added, alog which dislocations generated at the
crack tip can move away from the crack. Since bonds cannot be broken
and reformed in the linear zones, dislocations will be unable to leave
the non-linear zone.

The inter-atomic interactions in the non-linear zone are described by
the UBER pair potential:
\begin{equation}
  \label{eq:uber}
  F(r) = -k (r - r_0) \exp \left({r - r_0 \over \beta} \right)
\end{equation}
where $r$ is the inter-atomic separation, $r_0$ is the separation in
equilibrium, and $\beta$ is a range parameter.  We cut off the
interactions so that only nearest-neighbor interactions are included, and
shift the potential slightly to avoid a step in the force law.
Further, a small scaling of the force law is used to preserve the
elastic constants, thus enabling us to use the same Green's function
for all the force laws.  The force law thus becomes
\begin{equation}
  \label{eq:scaleduber}
  F(r) = C \left[ -k (r - r_0) \exp \left({r - r_0 \over \beta} \right) -
    F_0 \right]
\end{equation}
where $F_0 = -k (r_{\mbox{\scriptsize cutoff}} - r_0) \exp
\left((r_{\mbox{\scriptsize cutoff}} - r_0) / \beta \right)$ assures
that the force is zero at the cutoff distance $r_{\mbox{\scriptsize
    cutoff}}$ (in this work $1.7 r_0$).  This is only a slight
perturbation of the force law as the scaling factor $C$ only differs
from unity by a few percent.  Since this work does not study specific
materials, no attempt is made to use realistic many-body potentials.

\section{Results}

\begin{figure}[tbp]
  \begin{minipage}[t]{0.33\linewidth}
    \begin{center}
      a:\\
      \leavevmode
      \epsfig{file=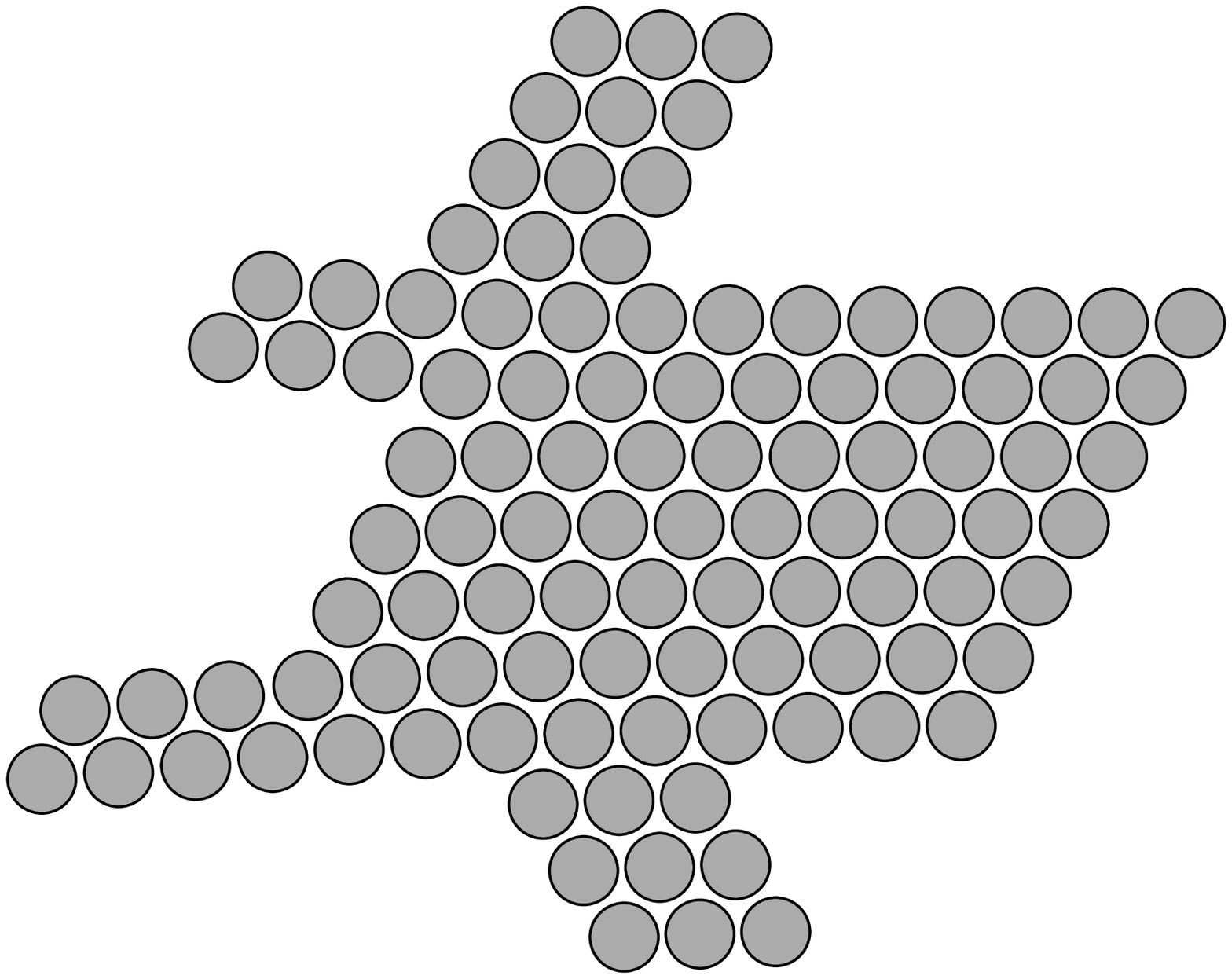,width=\linewidth}
    \end{center}
  \end{minipage}%
  \begin{minipage}[t]{0.33\linewidth}
    \begin{center}
      b:\\
      \leavevmode
      \epsfig{file=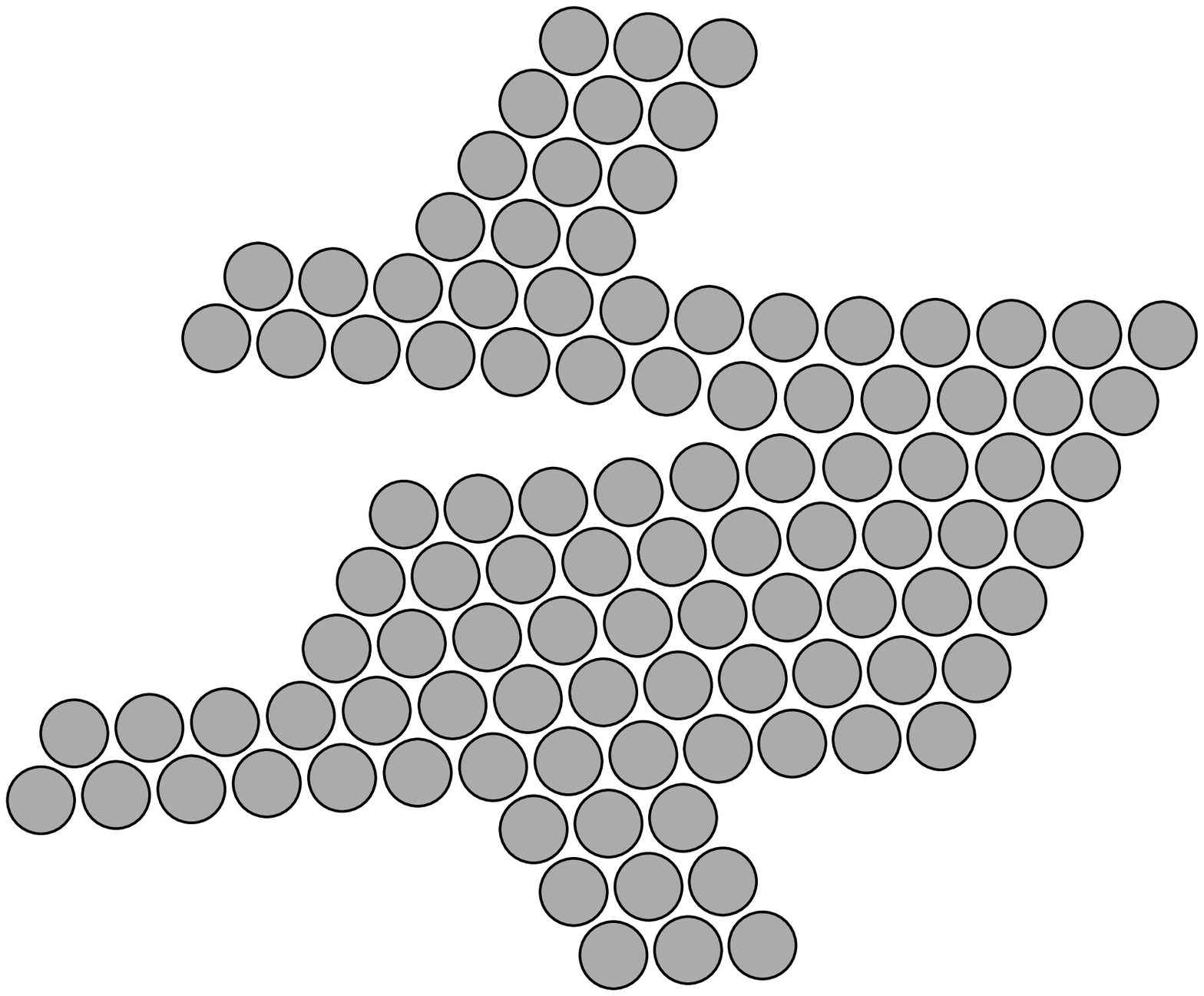,width=\linewidth}
    \end{center}
  \end{minipage}%
  \begin{minipage}[t]{0.33\linewidth}
    \begin{center}
      c:\\
      \leavevmode
      \epsfig{file=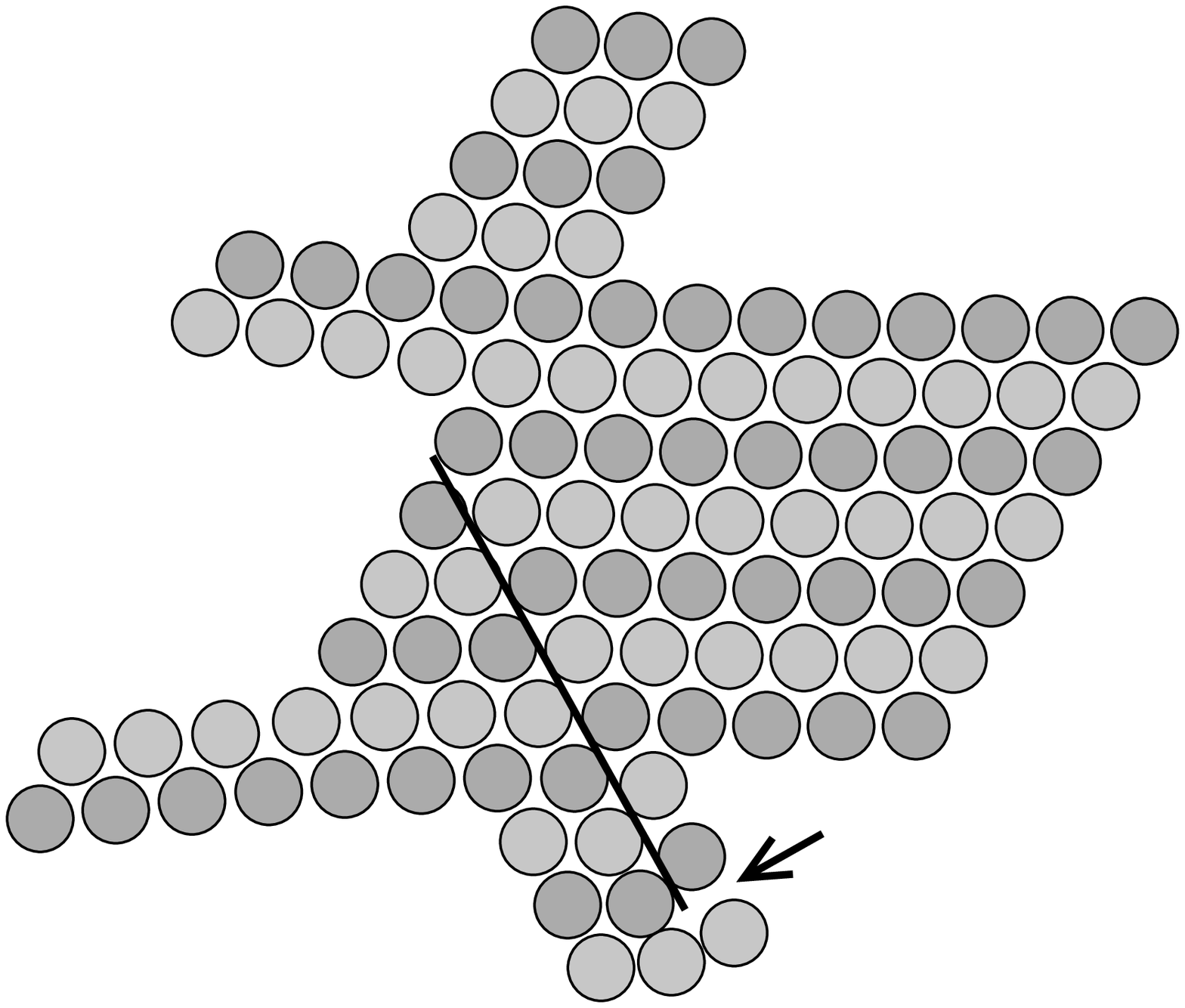,width=\linewidth}
    \end{center}
  \end{minipage}
  \caption[]{\footnotesize The two deformation modes.  (a) shows the
    configuration
    before plastic deformation.  The crack, three atomic layers thick,
    extends out of the left of the picture.  Only atoms in the
    non-linear zone are shown, see text.  (b) shows the crack
    propagating by cleavage.  (c) shows the crack emitting a
    dislocation.  The dislocation has traveled along the black line.
    Since it cannot leave the non-linear zone, it has been pinned at
    the bottom (indicated by the arrow).  To make the path of the
    dislocation visible, the atoms were given two different colors.
    Prior to emission the atoms were lined up in rows of the same
    color.  Where the dislocation has traveled lines of atoms of
    different color meet.}
  \label{fig:modes}
\end{figure}

When we load the cracks until plastic deformation occur, two different
deformation modes are observed: cleavage and dislocation emission, see
figure \ref{fig:modes}.  The dislocation emission always occurs in the
downwards direction, as shown in the figure, except for the case of a
single layer of blunting, where the crack geometry is symmetric.
Furthermore, the emission occurs in such a way that the asymmetric
crack geometry is preserved.  This means that further dislocations can
be expected to be emitted in the same direction, provided that the
dislocation can move so far away that its effect on the local stress
field at the crack tip is small.  When a sufficient number of
dislocations have been emitted, their combined screening may prevent
further emission, and cause the crack to propagate by cleavage or
begin emitting in the opposite direction.  This may have been observed
experimentally\cite{VeNe80,Oh85}.

\begin{figure}[tbp]
  \begin{center}
    \leavevmode
    \epsfig{file=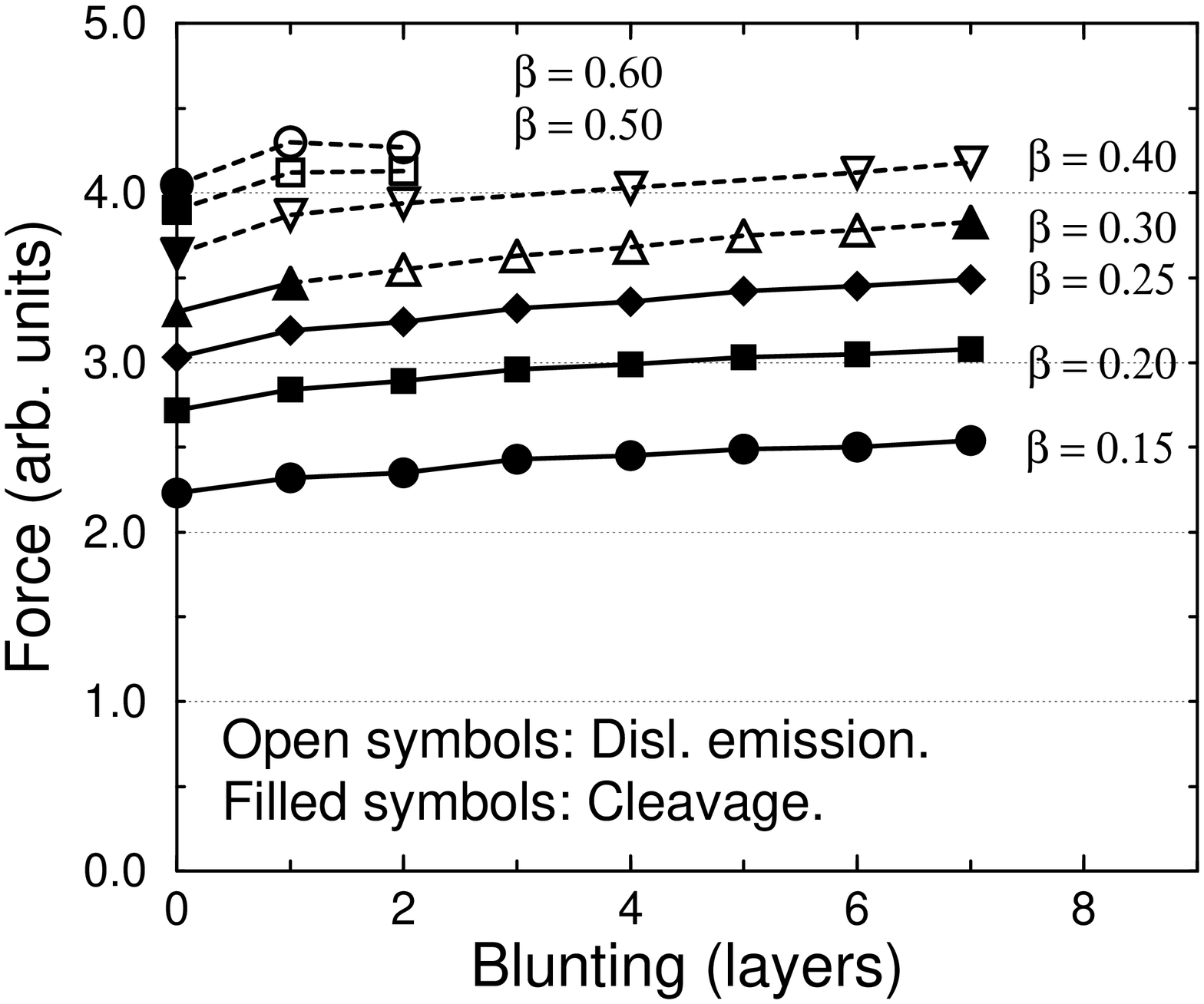,angle=-90,width=0.49\linewidth}%
    \epsfig{file=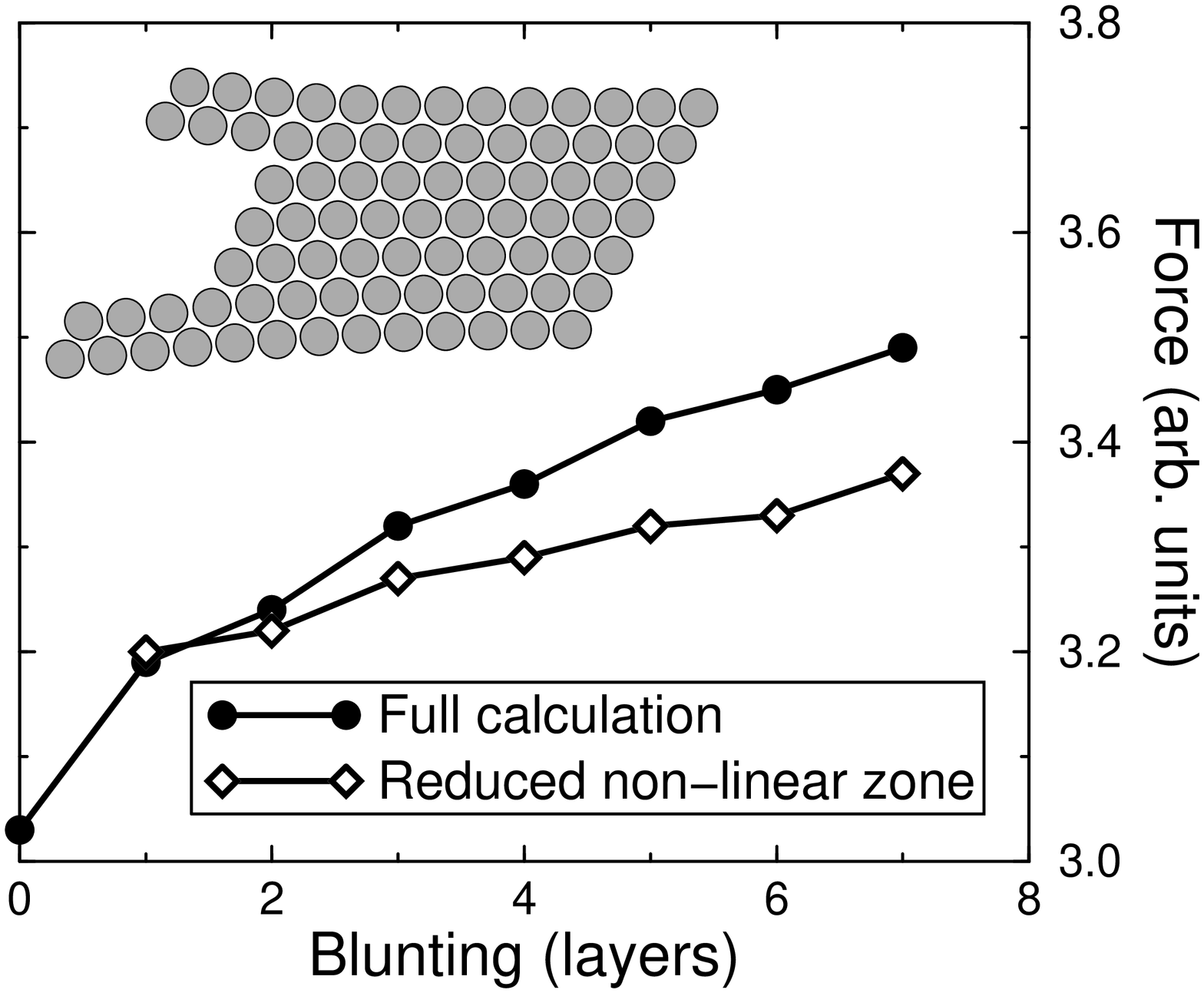,angle=-90,width=0.49\linewidth}
    \caption[]{\footnotesize Left: The force required for cleavage or
      dislocation emission as a function of the blunting, for seven
      different force laws.  Right: The effect of reducing the
      non-linear zone at the crack tip.  When the atoms shown in black
      are removed from the non-linear zone (i.e.\ they are restricted
      to interact through linear forces), the effect of the blunting
      is reduced.  The range parameter in the force law is $\beta =
      0.25$.}
    \label{fig:numresults}
  \end{center}
\end{figure}
Figure \ref{fig:numresults} (left) shows the force required for
cleavage or dislocation emission as a function of the blunting, for
seven different force laws.  All force laws result in cleavage in the
case of a sharp crack, but for a range of force law parameters
dislocation emission becomes favored as soon as the crack is blunted.
In all cases the required force for cleavage increases slightly with
increasing blunting.  When dislocation emission occur the force also
increases slightly with the blunting as soon as any blunting occurs,
but for these force laws the force required to emit a dislocation from
the sharp crack is {\em larger\/} that for the blunt crack.  The
tendency for the sharp crack to propagate is so large that it is
difficult to measure the dislocation emission criterion accurately.

Using a conformal mapping technique, we have been able to calculate
the stress field around a blunt crack (to be published elsewhere), but
only in anti-plane strain (mode III loading).  The result is as could
be expected: far from the crack the stress field is unperturbed by the
blunting, and decays as for the sharp crack ($\sigma \propto
r^{-1/2}$), but near the crack tip the stress singularity is similar
to what is found near a wedge-shaped crack\cite{Th86}:
\begin{equation}
  \label{eq:modeIIIwedge}
  \sigma \propto r^\alpha, \qquad \alpha = {\pi \over 2 \pi - \theta} - 1
\end{equation}
i.e.\ $\alpha = -2/5$ for the given configuration, a $60^\circ$ wedge,
as opposed to $-1/2$ for a sharp crack ($\theta = 0$).
This leads to a reduction of the stresses near the crack tip, and thus
to an increase in the crack loading needed to break the bonds.
However, this leads us to expect an increase in the load of
approximately 25--50\% over the range of blunting investigated here,
versus the observed increase of only $\sim 10\%$.  This is caused by
the difference between the mode I and mode III configurations.
Williams\cite{Wi52} has solved the mode I wedge configuration and the
stress singularity is significantly different.  He finds $\alpha =
-0.478$ for this geometry, leading to a stress singularity that is
indistinguishable from the sharp crack.  This would lead us to expect
that there should be no effect from the blunting.  So we have to seek the
explanation of the observed effect elsewhere than in linear
elasticity.

Figure \ref{fig:numresults} (right) shows the result of reducing the
number of atoms interacting through non-linear forces.  This strongly
suppresses the effect of the blunting, indicating that it is mostly a
non-linear effect.  When the crack is loaded the bonds at the blunt end of
the crack are stretched into their non-linear regime.  They are thus
stretched more than they would have been if the bonds had been purely
linear, and the bonds at the corner where the crack appear do not have
to stretch as much, and therefore do not break as early.  As the
blunting increases, more non-linear bonds become available for
relieving the stress concentration at the corner.

\section{Conclusions}

We have shown that blunt cracks have a stronger tendency to emit
dislocations than do sharp cracks.  This may be important in
intrinsically brittle materials, where the natural tendency for a
sharp crack is to propagate by cleavage rather than to emit
dislocations.  In these materials a dislocation may be blunted by
{\em absorbing\/} a dislocation from a nearby source, and thereby be
turned into an emitting crack, preventing further cleavage.

We also observe, that dislocation emission from an asymmetric blunt
crack preferentially occurs to one side, and that the emission
preserves the shape of the crack.  This leads to emission of multiple
dislocations to the same side, even in situations where emission on
two symmetrical slip planes should be equally favorable, and where a
simple shielding argument would result in emission of every second
dislocation on alternate planes.

\section{Acknowledgments}

The authors would like to thank Rob Phillips for many useful
discussions.  This work was in part supported by the National Institute
of Standards and Technology under award 60NANB4D1587, and by the
Office of Naval Research under Grant Nos.\ N00014-92-J-4049 and
N00014-92-F-0098.


\end{document}